\documentstyle[twocolumn,prc,aps]{revtex}
\newcommand{ \be }{\begin{equation}}
\newcommand{ \ee }{\end{equation}}
\newcommand{ \bea }{\begin{eqnarray}}
\newcommand{ \eea }{\end{eqnarray}}
\newcommand{ \la }{\langle}
\newcommand{ \ra }{\rangle}
\newcommand{ \bp }{{\bf p}}

\newcommand{ \ds }{\displaystyle}

\begin{document}
\draft

\title{ On the normalization of the HBT correlation function}

\author{
  D.~Mi\'{s}kowiec$^1$ and
  S.~Voloshin$^{2,}$\thanks{On leave from Moscow Engineering Physics 
   			    Institute, Moscow, 115409, Russia. }}
\address{
 $^1$ Gesellschaft f\"ur Schwerionenforschung, Darmstadt, Germany\\
 $^2$ Physikalisches Institut der Universit\"at Heidelberg, Germany}

\date{\today}

\maketitle

\begin{abstract}
We discuss the question of the normalization of the correlation
function and  its consistency with the often used form
$ C(\bp_1,\bp_2) = 1+|f(\bp_1,\bp_2)|^2 $.
We suggest an event mixing method which allows one to obtain absolutely 
normalized correlation functions from experimental data. 
\end{abstract}


In the current note we address a question which being simple has
caused confusing and erroneous statements in the literature.
We discuss the question of the normalization of the
correlation function used in the HBT analysis of multiparticle
production.
The normalization is, of course, a matter of convenience and author's
favor.
However, the question of the normalization becomes important
if one wants to compare the results from different papers or to compare
the experimental data to the theoretical predictions.
In this note we show that in the literature there exist basically two
definitions of the correlation function which differ by the normalization.
Remarkably, starting from these two different definitions,
authors arrive to the same result for the relation between
the correlation  function and the particle emission (source) function.
This means that some of the derivations are wrong.


The different definitions of the  two-particle correlation function
used in the theoretical papers on two-particle correlations
in heavy ion collisions are the following. 
The first definition~\cite{yan78,ber94,gyu88,pad90,mak88,ber88,mro95} is
\be
\label{c1}
C_1(\bp_1,\bp_2)  =
\frac{\ds
\sigma \frac{d^6\sigma}{d^3p_1 d^3p_2}
}{\ds
\frac{d^3\sigma}{d^3p_1} \frac{d^3\sigma}{d^3p_2}} =
\frac{\ds
\frac{d^6n}{d^3p_1 d^3p_2}
}{\ds
\frac{d^3n}{d^3p_1} \frac{d^3n}{d^3p_2}} \ ,\\
\ee
where $d^3\sigma/d^3p$ and $d^6\sigma/d^3p_1/d^3p_2$
are the single-particle and the two-particle
cross sections, respectively,
and $\sigma$ is the total (inelastic) cross section,
while
$d^3n/d^3p$ and $d^6n/d^3p_1/d^3p_2$ are the particle and the
pair yields per event, respectively.
The relations between the cross sections and yields are
$d^3n/d^3p=1/\sigma \ d^3\sigma/d^3p$,
$d^6n/d^3p_1/d^3p_2=1/\sigma \ d^6\sigma/d^3p_1 /d^3p_2$.

The second definition~ \cite{gyu79,cso96,cha94,wie96}
differs from the first one by the normalization factor:
\be
\label{c2}
C_2(\bp_1,\bp_2) =
\frac{\langle n\rangle^2}{\langle n(n-1)\rangle}
\frac{\ds
\frac{d^6n}{d^3p_1 d^3p_2}
}{\ds
\frac{d^3n}{d^3p_1} \frac{d^3n}{d^3p_2}}
\ee
with $\langle n\rangle$ and $\langle n(n-1)\rangle$ being the mean
number of particles and mean number of particle pairs per event,
respectively.

The third definition~\cite{shu73,pra84,pra86,pra86b,pra90,her95} is
\be
\label{c3}
C_3(\bp_1,\bp_2) =
\frac{P({\bp_1},{\bp_2})}{P({\bp_1}) P({\bp_2})},
\ee
where $P({\bp_1})$ and $P({\bp_1},{\bp_2})$ are called
probabilities of observing a single particle with momentum $\bp_1$
and a pair of particles with momenta $\bp_1$ and $\bp_2$, respectively.
Such a definition is not strict and may cause confusion.
The word `probability' can be understood in two ways: 
1) $P({\bp_1}) d{\bp_1}$ is the probability of observing in a given event
a particle with momentum between $\bp_1$ and $\bp_1+d\bp_1$ ; 
in this case $P({\bp_1})=d^3n/d^3p_1$,
or 2) $P({\bp_1})$ is the probability density for a given particle to 
have momentum $\bp_1$.
In the last case
$P({\bp_1})=1/\la n \ra \ d^3n/d^3p_1$.
Thus, depending on the interpretation of
$P({\bp_1})$ and $P({\bp_1},{\bp_2})$,
the definition (\ref{c3}) becomes equivalent to
(\ref{c1}) or (\ref{c2}).
In the rest of this note we consider the difference between
(\ref{c1}) and (\ref{c2}).

The factor, by which the definition (\ref{c2}) differs from (\ref{c1}), is
usually close to unity. For a Poissonian multiplicity distribution it is
exactly unity. In the presence of two-particle correlations, however, one
cannot assume that the multiplicity distribution is Poissonian.
Thus, in a general situation, the two definitions are different.
Still, using these two different normalizations, different authors
arrive to the same expression relating the correlation function
to the source function.
Brief review of the literature shows that the reason for this is that,
in spite of declaring different normalizations at the beginning,
everybody (except, probably, \cite{gyu79} where the normalization is kept
through-out the paper) starts his/her calculation from the same
expression for the correlation function in terms of
the creation and annihilation operators (or analogous one in terms of
$\cal T$-matrix):
\be
\label{ca}
C=\frac{\la a^\dagger_1 a^\dagger_2 a_2 a_1 \ra}
{\la a^\dagger_1 a_1 \ra \la a^\dagger_2 a_2 \ra} \ .
\ee
This equation, for chaotic source and Bose-Einstein statistics,
leads (see for example~\cite{chao94}) to the correlation function in the form:
\be
\label{ct}
C(\bp_1,\bp_2) = 1+\frac{|\la a^\dagger_1 a_2 \ra |^2}
{\la a^\dagger_1 a_1 \ra \la a^\dagger_2 a_2 \ra}=
1+|f(\bp_1,\bp_2)|^2 \,
\ee
with the function $f(\bp_1,\bp_2)$ depending on the shape of the boson source
and asymptotically approaching zero for $|\bp_2-\bp_1|\rightarrow \infty$.
For the Fermi-Dirac statistics there is a minus sign in front
of the second term.


The equivalence of the definitions (\ref{c1}) and (\ref{ca}) can easily
be shown using the expressions for one and two particle invariant
distributions:
\be
E_1\frac{d^3n}{d^3p_1}=\la a^\dagger_1 a_1 \ra,
\,\,\, \,\,\, \,\,\,
E_1 E_2 \frac{d^6n}{d^3p_1 d^3p_2}
=\la a^\dagger_1 a^\dagger_2 a_2 a_1 \ra \ .
\ee


It also can be shown that the definition~(\ref{c2}) and Eq.~(\ref{ct}) are
inconsistent.
Indeed, combining them we obtain
\be
\label{wrong}
\frac{d^6n}{d^3p_1 d^3p_2}
=
\frac{\langle n(n-1)\rangle}{\langle n\rangle^2} \
\frac{d^3n}{d^3p_1} \frac{d^3n}{d^3p_2} \
\left(1+|f(\bp_1,\bp_2)|^2\right)\ .\\
\ee
Let us integrate this equation over ${\bp_1}$ and ${\bp_2}$.
Since the integral of $d^6n/d^3p_1 /d^3p_2$ is equal to the
mean number of pairs per event $\langle n(n-1)\rangle$,
and the integral of $d^3n/d^3p$ is the mean multiplicity
$\langle n\rangle$, the integration of (\ref{wrong}) leads to
\be
\int{\frac{d^3n}{d^3p_1} \frac{d^3n}{d^3p_2}} \ |f(\bp_1,\bp_2)|^2 \
d^3p_1 d^3p_2 = 0,
\ee
which is not possible unless $f(\bp_1,\bp_2)$ is identically equal to zero.
Thus, the derivation starting from (\ref{c2}) and leading
to (\ref{ct}) must contain an error.

An expression analogous to (\ref{wrong}) but obtained using the definition
(\ref{c1}) would not contain the factor
$\langle n(n-1)\rangle / \langle n\rangle^2$.
An integration of this equation in absence of correlations gives
$\la n(n-1) \ra  = \la n \ra^2$.
For identical bosons (without final state interactions)
the correlation function is greater than unity,
and thus the same integration yields $\la n(n-1) \ra  > \la n \ra^2$
which reflects the fact that bosons like to be produced copiously.
For fermions the corresponding correlation function $C(\bp_1,\bp_2)<1$.
This results in $\la n(n-1) \ra  < \la n \ra^2$ which reflects Pauli blocking.


The whole presented problem is meaningless from the practical point of 
view as far as the experimentalists
do not use Eq.~(\ref{c1}) nor~(\ref{c2}) to normalize the measured
correlation functions.
Instead, based on (\ref{ct}), they choose the normalization such that
$C(\bp_1,\bp_2) \rightarrow 1$ for $|\bp_2-\bp_1| \rightarrow \infty$.
However, after we established that the definition (\ref{c1}) is correct
in the sense that it matches (\ref{ct}),
we may note that it does not leave any freedom of the normalization.
For the often used event mixing technique it would mean the following.
As usual, to generate mixed pairs one should take a pair of different 
events and combine all particles from one event with all particles 
from the other event. 
The normalization given by (\ref{c1}) will be automatically achieved 
if the number of event pairs used in event mixing is equal to the number of
events used to calculate the actual pair distribution.  

Summarizing, we showed that the correlation function in the form~(\ref{ct}), 
often used in theoretical and experimental papers, 
corresponds to the definition~(\ref{c1}) and is not consistent with
definition~(\ref{c2}). 
We suggest to use the definition~(\ref{c1}) as the basis 
for the analysis of experimental data.

We thank P.~Braun-Munzinger, T.~Cs\"org\H{o}, U.~Heinz, S.~Pratt, 
and W.~Zajc for discussions.


\begin{thebibliography}{99}
\bibitem{yan78}
F.B.~Yano and S.E.~Koonin, Phys. Lett. 78 B, 556 (1978).
\bibitem{ber94}
G.F.~Bertsch, P.~Danielewicz, and M.~Herrmann, Phys. Rev. C 49, 442 (1994).
\bibitem{gyu88}
M.~Gyulassy and S.~Padula, Phys. Lett. B 217(1988)181.
\bibitem{pad90}
S.S.~Padula, M.~Gyulassy, and S.~Gavin, Nucl.Phys.B329(1990)357.
\bibitem{mak88}
A.N.~Makhlin and Yu.M.~Sinyukov, Z.Phys.C 39,69 (1988).
\bibitem{ber88}
G.~Bertsch, M.~Gong, and M.~Tohyama, Phys. Rev. C 37, 1896 (1988).
\bibitem{mro95}
S.~Mr\'owczy\'nski, Phys. Lett. B 345(1995)393.
\bibitem{gyu79}
M.~Gyulassy, S.K.~Kauffmann, and L.~Wilson, Phys. Rev. C 20(1979)2267.
\bibitem{cso96}
T.~Cs\"org\H{o}, B.~L\"orstad, and J.~Zim\'anyi, Z.Phys. C 71(1996)491.
\bibitem{cha94}
S.~Chapman and U.~Heinz, Phys.~Lett. B 340, 250 (1994).
\bibitem{wie96}
U.A.~Wiedemann, P.~Scotto, and U.~Heinz, Phys.Rev.C 53, 918 (1996).
\bibitem{shu73}
E.V.~Shuryak, Phys. Lett. 44 B (1973) 387.
\bibitem{pra84}
S.~Pratt, Phys. Rev. Lett. 53, 1219 (1984).
\bibitem{pra86}
S.~Pratt, Phys. Rev. D 33(1986)72.
\bibitem{pra86b}
S.~Pratt, Phys. Rev. D 33(1986)1314.
\bibitem{pra90}
S.~Pratt, Phys. Rev. C 42(1990)2646.
\bibitem{her95}
M.~Herrmann and G.F.~Bertsch, Phys. Rev. C 51 (1995) 328.
\bibitem{chao94}
W.Q.~Chao, C.S.~Gao, and Q.H.~Zhang, Phys. Rev. C 49 (1994) 3224.
\end{thebibliography}
\end{document}